\documentclass[3p,times,twocolumn]{elsarticle}

\usepackage{ecrc}

\volume{00}

\firstpage{1}

\journalname{Nuclear Physics B Proceedings Supplement}

\runauth{D. Fargion et al.}

\jid{nuphbp}

\jnltitlelogo{Nuclear Physics B Proceedings Supplement}

\usepackage{amssymb}
\usepackage{color}

\usepackage[figuresright]{rotating}
\usepackage{enumitem}
\usepackage[utf8]{inputenc}
\usepackage{hyperref}

\begin{document}

\begin{frontmatter}

\title{UHECR narrow clustering correlating IceCube through-going muons}

\author[label0,label1,label3]{Daniele Fargion}
 \ead{daniele.fargion@roma1.infn.it}
 \author[label2,label3]{Pietro Oliva}
 \author[label4]{Pier Giorgio De Sanctis Lucentini}
\author[]{Daniele D'Armiento}
\author[label0]{Paolo Paggi}

  \address[label0]{Physics Department, Rome University 1, P.le A. Moro 2, 00185, Rome, Italy}
 \address[label1]{INFN Rome1, P.le A. Moro 2, 00185, Rome, Italy}
  \address[label2]{Department of Electrical Engineering, Niccol\`o Cusano University, Via Don Carlo Gnocchi 3, 00166 Rome, Italy}
 \address[label3]{MIFP, Mediterranean Institute of Fundamental Physics - Via Appia Nuova 31, 00040 Marino (Rome), Italy}
\address[label4]{Gubkin Russian State University
of Oil and Gas (National Research University), 65 Leninsky prospect, 119991 Moscow, Russian Federation.}

\begin{abstract}
The recent UHECR events by AUGER and the Telescope Array (TA) suggested that wide clusterings as the North and South, named Hot Spot, are related to near AGNs such as the one in M82 and Cen~A (as most authors and us convened). In the same frame since 2008 we assumed that the UHECR are made by light and lightest nuclei to explain the otherwise embarrassing absence of the huge nearby Virgo cluster, absence due to the fragility and the opacity of lightest nuclei by photo-dissociation from Virgo distances. Moreover UHECR map exhibits a few narrow clustering, some near the galactic plane, as toward SS~433 and on the opposite side of the plane at celestial horizons: we tagged them in 2014 suggesting possible near source active also as a UHE neutrino. Indeed since last year, 2015, highest IceCube  trough-going muons, UHE up-going neutrino events at hundreds TeV energy,
 did show (by two cases over three tagged in North sky) the expected overlapping of UHE neutrinos signals with narrow crowding UHECR.
  New recent data with higher energy threshold somehow re-confirmed our preliminary proposal offering also new possible
  sources by a additional correlated UHE-neutrino versus UHE-neutrino and-or with narrow UHECR clustering events.

\end{abstract}

\begin{keyword}
Cosmic Rays \sep Neutrino Astronomy

\end{keyword}

\end{frontmatter}

\section{2013-2016: IceCube (missing) astronomy}

For three years now, since November 2013, IceCube claimed the discover of astrophysical UHE neutrinos~\cite{007},
but not the birth of UHE neutrino astronomy which is yet to come.
Since nine years, November 2007, the Pierre Auger Collaboration claimed a definitive connection between the UHECR and the nearby (local) group within the expected GZK cut off \cite{0a, 0b}; however, this opening for a new particle astronomy~\cite{006} failed with more data and increased statistics, leaving us today with only a mild twin hot spot anisotropy in the UHECR sky \cite{19}, or even worse with just a meaningless spread homogeneous sky \cite{1}. We claimed, in lonely corner, since 2008 that these spread Hot Spot are made by lightest UHECR nuclei partially bent and almost totally absorbed from Virgo cluster. This may explain its UHECR embarrassing absence.
The above mentioned astrophysical neutrino discover at hundred TeV-PeV energy on 2013 was mainly based on a tiny hardening of the neutrino spectra around 30~TeV with respect to the otherwise dominant atmospheric high energy muon neutrino tracks. In particular it was based on the overproduction of showering events (nearly spherical lightnings in diffused ice) called cascades. This cascades abundance over muon tracks underline a sudden neutrino flavor change.
The most popular and acclaimed interpretations still widely accepted were that
above 30~TeV, or in some cases over 60~TeV, most of the IceCube neutrino events are of astrophysical nature.

The flavor change has been the most compelling and convincing signature
of a new component in IceCube signals \cite{27}. Indeed, at tens TeV, the atmospheric neutrinos are still ruled by the muon ones by a factor of almost twenty (respect to electron ones) because the primary ultra-relativistic muons cannot decay easily inside Earth's short atmosphere (into electrons) above tens-hundred GeV energies, while shorter life pion and the kaons may still decay in flight into muons.
 On the contrary, any  primordial astrophysical neutrinos, either muon or electron at birth, would oscillate
over the long interstellar distances and they will result into a democratic comparable flavor ratio $\frac{1}{3}:\frac{1}{3}:\frac{1}{3}$ \cite{0e1}, no more ruled by muon ones but  by shower (cascade) events.
 Because of it, one would expect that the UHE astrophysical neutrino showering events, mostly in debt to astrophysical electron, or tau, or neutral current neutrino interaction with IceCube nuclei, would lead to IceCube
 showers (cascades) approximately $3/4$  over all the other signals (both shower or muon tracks).
Indeed, the first 37 events above 30 TeV in IceCube  during 2013-2014 as well more recent 54 ones did show
 such a ruling shower flavor predominance over the more diluted and rare muon tracks.
However, there are several missing pieces of the puzzle:\newline
a) Why not yet any clustering of these UHE$\nu$ toward expected highest energy $\gamma$~sources?\newline
b) Why lower energetic (TeVs) neutrino events are
 homogeneous in the sky?\newline
c) Why GRBs are not correlated with any of such UHE$\nu$ events?\newline
d) Why High Energy neutrinos are not (apparently) along the galactic plane?\newline
e) Why within a dozen of 200 TeV $\nu$ event the $\nu_\tau$ didn't rise (yet)?
\begin{figure}[t]
\centering
\includegraphics[width = 0.90\columnwidth]{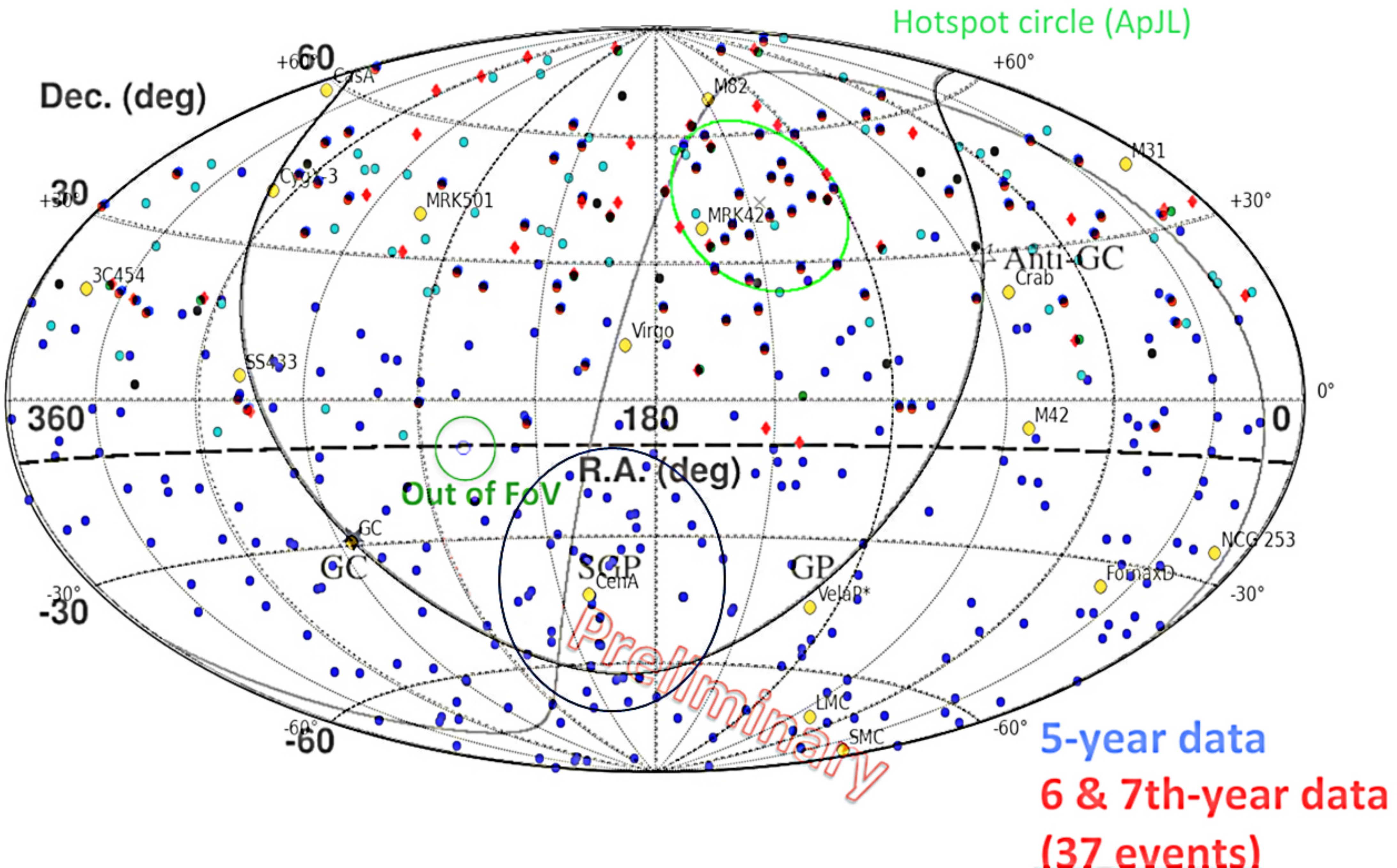}
\caption{The two Hot Spot in UHECR and the two candidate sources by M82 (North) and Cen~A (South) of the celestial coordinate sky. Blue events are by AUGER while the red ones by TA UHECR with updated last (6 and 7) years events (2016); the magenta dots stand for the old AGASA events}\label{fig:1}
\end{figure}

 Some possible answers have been offered only recently \cite{19} 
  and mentioned below.
The rare and eventually fortuitous correlation between one of highest energy IceCube cascades and an AGN flare activity \cite{26} is not solving the puzzles above. Indeed the absence of GRB neutrino might find an answer in the GRB model itself
being with no hadronic origination but just by a pure electronic pair jet  \cite{23}.
\begin{figure}[t]
\centering
\includegraphics[width = 0.90\columnwidth]{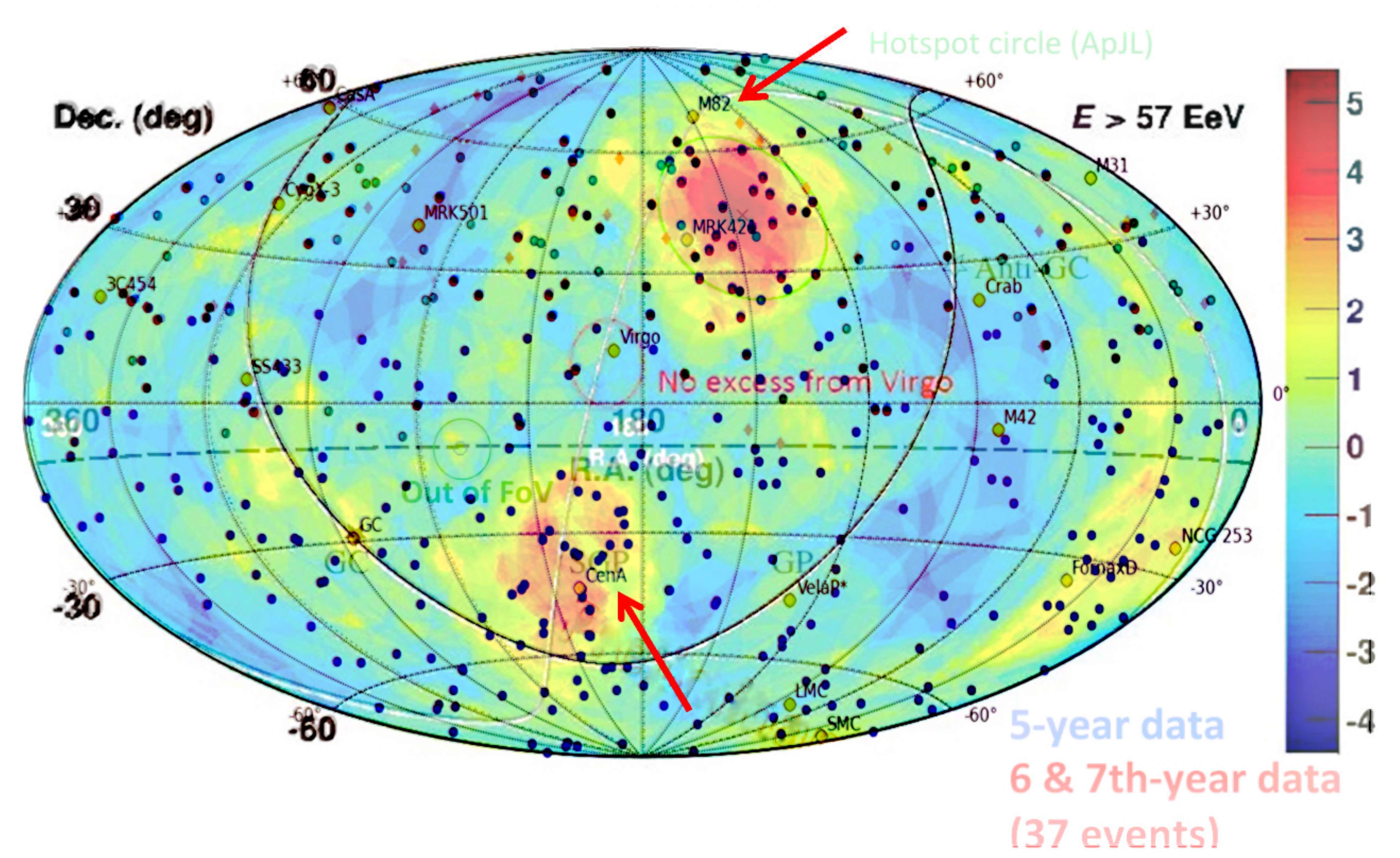}
\caption{As above the two Hot Spot in UHECR and the two candidate sources by M82 (north) and Cen A (south) of the celestial coordinate sky; note also some additional nearby source as the LMC and SMC, Fornax Galaxy, NGC 253 near AGN; note also nearby galactic narrow clustering toward remarkable nearby Vela and Cygnus X3 gamma sources, as well toward  SS433 where cluster C raised.}\label{fig:2}
\end{figure}

\begin{figure}[t]
\centering
\includegraphics[width = 0.90\columnwidth]{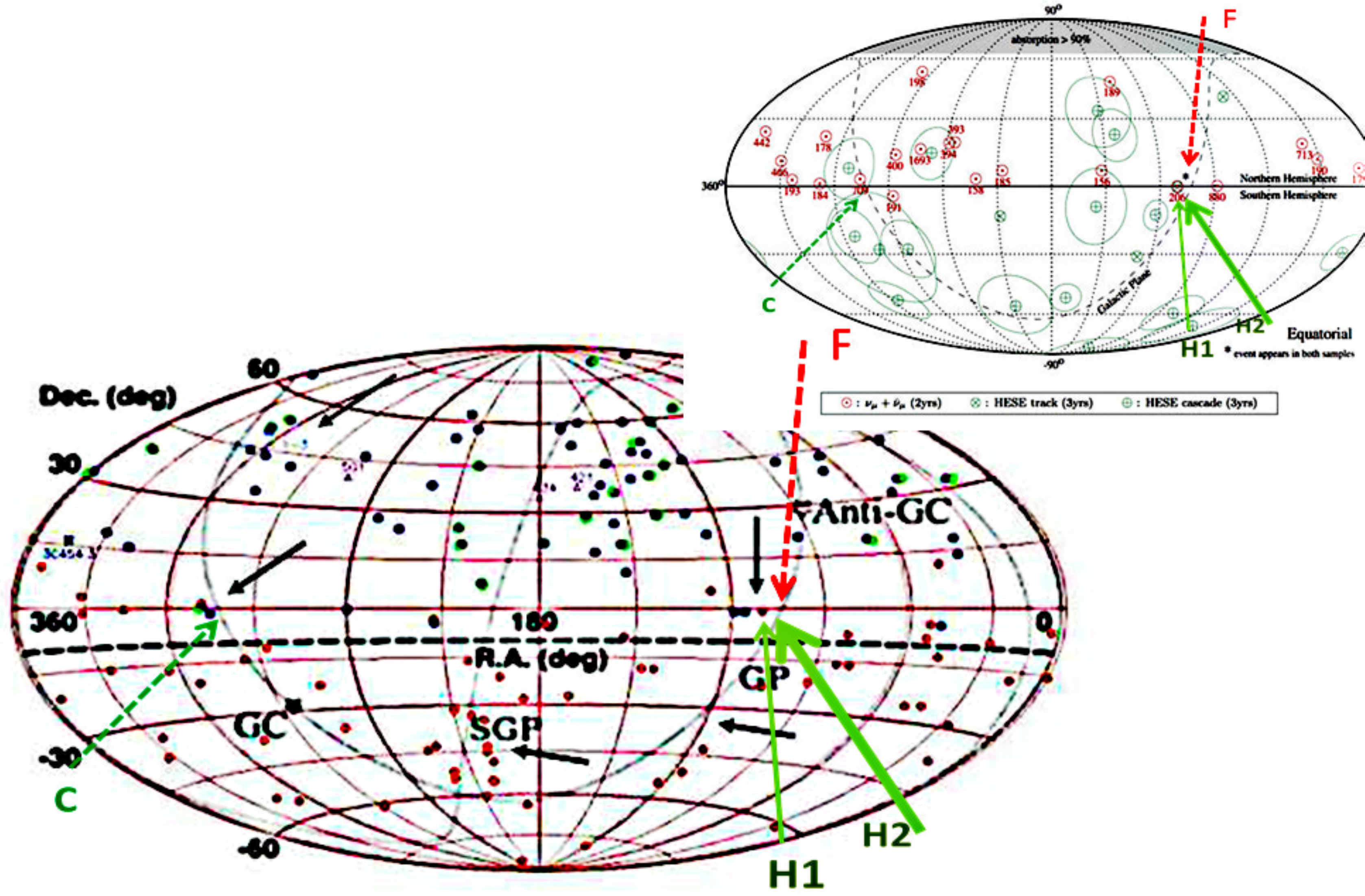}
 \caption{A representation of the Sky map (2014)  in equatorial coordinates with the early UHECR clustering events 2013 whose
 narrow ones, tagged by black arrows, were suggesting possible  UHE correlated neutrinos \cite{8a}. Among the three (of five) black arrow in the North sky, two directions, C and (H1),have been later on (2015) indeed observed as the crossing muon tracks. The up figures show the arrival direction of the 21 highest-energy events (2015) trough-going muons at energy above 150 TeV. The most probable neutrino energy (in TeV) indicated for each event assumes the best-fit astrophysical flux of the analysis \cite{6e}. For comparison, the events of the 3-year high-energy starting event (HESE) analysis with deposited energy larger than 60 TeV (tracks and cascades) are also shown. As we mention we believe that mostly hundreds TeV energy threshold select astrophysical events. Cascade events are indicated together with their median angular uncertainty (thin circles); some of those rare UHECR event toward SS 433, label C, has been note in earliest time \cite{12} as in case of cluster H  respectively observed on 2015 (H1) and 2016 (H2, F). See table \ref{table:1}.}
	\label{fig:3}
\end{figure}
 However, the same a), b), c), d), e) questions might find an answer thanks to a different (additional o hybrid)
     neutrino component, able to change the flavor ratio components:
     the charmed neutrinos, whose spectra might mimic the Cosmic Ray one with an exponent
     around the observed $2.2-2.7$. The injection of a dominant atmospheric prompt neutrino
     or the sudden change in chemical composition of the CR at the knee
     might explain at once most of all the above questions: no clustering to expected sources
     (as their parental CR), no self clustering (as most CR ones), obviously no GRB connection, no galactic plane signature
     nor any $\tau$ double bang \cite{0c}. In this pessimistic ``all-charmed'' neutrino signals in IceCube,
      the atmospheric tau charmed neutrinos should nevertheless rise \cite{19} 
     above 200 TeV as soon as nearly a dozen of muon high-energy starting event (HESE) tracks will be recorded;
     this must occur anyway because the tau appearance in prompt neutrinos is nearly one-tenth
     of the muon cross section. This imply such a tau event in a wider record of three dozen of events above 200 TeV
     possibly in the future 8 years or more. In a more optimistic view the astrophysical
     neutrino may be the key desired additional  component of the whole IceCube data set above several hundred TeV.
    In this view a probable tau double bang, due to real astrophysical neutrino tau, must rise very soon
   within, let's say, a couple of years offering the clear imprint mark of astrophysics source.
   In the same view the up-going UHE muon tracks originated outside the same IceCube detector maybe
   mostly or in majority of astrophysical nature.

 \section{IceCube neutrinos and through-going muons}
\begin{figure}[t]
\centering
\includegraphics[width = 0.90\columnwidth]{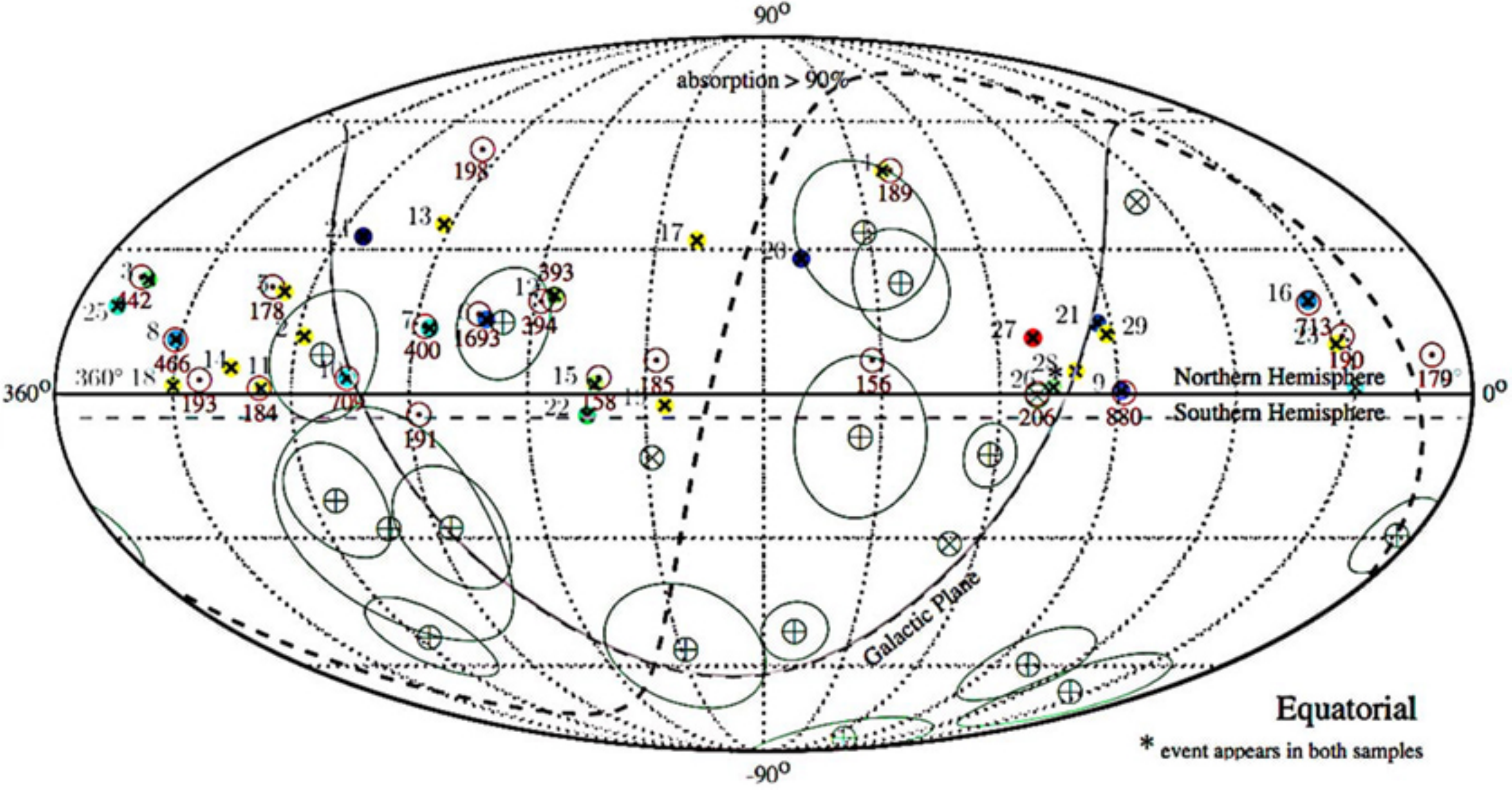}
\caption{The 2015 and 2016 through-going horizontal and upwards in celestial coordinate. It should be noticed that out of the recent 29 events, a large fraction are the same ones of 2015 with minor changes; 2015 low energy events have been filtered.}
\label{fig:4}
\end{figure}

   Indeed the first three PeV energetic showers (cascades) have been united to a novel
   highest energetic through-going muons track  at 2.6 PeV (energy released) whose
   discover on $11$ June, 2014 has been published just a few months ago \cite{20}; this event and the additional
   trough going ones above 150 TeVs have been published (in the IceCube web site \cite{6e}) last year, October 2015, and
    somehow updated with different data and higher energy threshold recently \cite{21};  they are shown in 2015 IceCube map, see Fig.~\ref{fig:3}, and in the last one of 2016, see Fig.~\ref{fig:4}. The first UHECR narrow clustering and its possible correlation have been discussed in a recent work and in the figures and references therein \cite{22}.
As soon as the IceCube PeV energy flavor neutrino revolution had occurred and has been analyzed \cite{27}, the smeared directionality and the useless correlation of the shower (cascades) events as well as the paucity of highest energy muon neutrino tracks forced us to imagine a very different neutrino astronomies: the crossing or through-going $\mu$ ones \cite{0c}.  Indeed, we claimed the urgency of an horizontal-upgoing muon neutrino astronomy \cite{0c} because of the longevity and the length of a PeV muon track, because of their consequent much wider effective detection volume (not confined as for the shower -cascades- ones inside the km$^3$ detector)  and finally because of their considerably better directionality.  Only recently such a $\mu$ track records where considered (in 2015 \cite{6e}) and a more recent IceCube study updated it and focused the attention on the up-going muon signals \cite{21}.
The idea that mainly upward crossing muons (through-going) are the cornerstone in IceCube neutrino astronomy \cite{0c}, rose as soon as the dominant showering events (cascades) did show a spread ($\pm 10^{\circ}$) arrival directions, in few dozen events in IceCube above 60 TeV energy \cite{007}.
Only a small fraction of those tracks offered a sharp directionality useful for astronomical correlations: the events of the 3-year high-energy starting event -HESE- analysis with deposited energy larger than 60 TeV, tracks and cascades, shown in Fig.~\ref{fig:1}.
The UHE crossing muons event labeled as, $\textbf{1}$-$\textbf{29}$, as in \cite{21} are not totally identical to the previous year ones. Firstly, there is a higher energy threshold (in 2015 it was 150 TeV, in 2016 it is above 200 TeV), secondly several event are in different arrival angle and energy. Thirdly, as a consequence, the direction changes hint for a poorer angular resolution even for muon tracks. Thus,  one may relax the angular correlation as sharp as the expected $\pm 0.4^{\circ}$ to a wider empirical  $\pm 3.7^{\circ}$.
This has some remarkable consequences in the statistical self-correlation of the through-going muon track events and on the possible pairs or triplet events. Let us comment the 2015-2016 through-going muon event from the largest right ascension to the smallest ones (from the left to the right of Fig.~\ref{fig:4}).

\subsection{Events in detail and Table reading}
\label{Events in detail and Table reading}
\begin{figure}[t]
\centering
\includegraphics[width = 0.90\columnwidth]{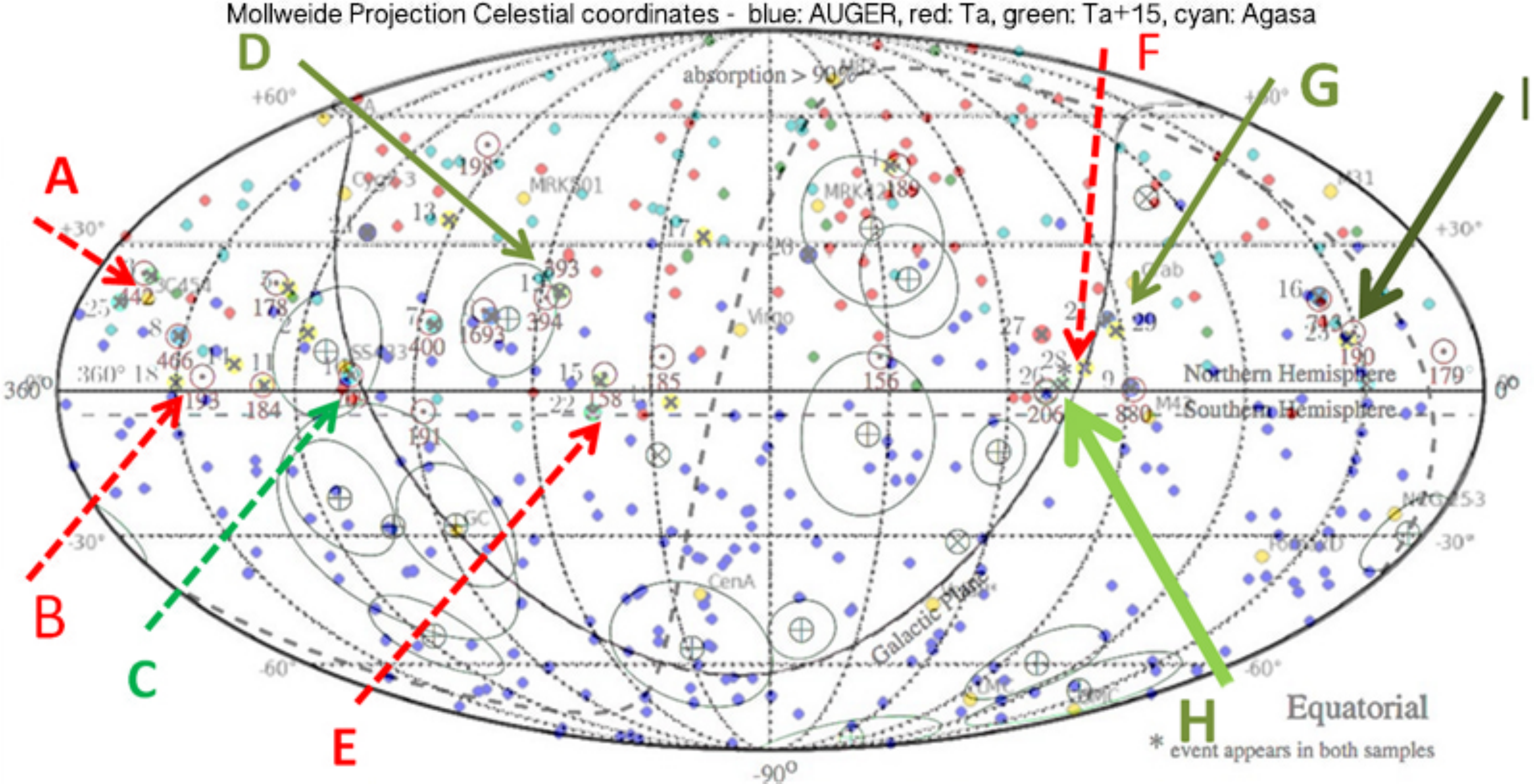}
\caption{The 2015 and 2016 through-going horizontal and upwards in celestial coordinate. A large fraction are the same ones of 2015 with minor changes; nine low energy events have been filtered.}
\label{fig:5}
\end{figure}

The new 29 through-going muon tracks (2016) versus the earlier catalog ones (2015)
present 12 overlapped events (let call them, coincident) in a somehow hidden form here below explained.
Most of them show a different estimate of the parental UHE $\nu_\mu$ energy,
 while a large fraction of them exhibit even a different arrival direction shifted up to $\simeq 3.7^{\circ}$ away.
 In particular: New event 3 is coincident, but less energetic, with earlier 442 TeV event. New event 8 is coincident, but more energetic, with earlier 466 TeV event. New event 11 is coincident and more energetic than earlier event of 184 TeV.
New event 5 is coincident, but more energetic and widely displaced than earlier event 178 TeV.
New event 10 is coincident but less energetic than event 709 TeV.New event 7 is coincident and less energetic of event label by 400 TeV. New event 6 is coincident but less energetic and displaced than earlier label 1693 TeV. New event 12 is coincident but less energetic than earlier one label 393 TeV. New event 15 is coincident but twice energetic than earlier event label 158 TeV. New event 4 in coincident but displaced than earlier label event 189 TeV. New event 26 is not coincident with earlier one label by 206 TeV through-going as well as coincident to a inner HESE event n.5. New event 9 is coincident and more energetic than earlier one label by 880 TeV.  New event 21 and New event 29 are not coincident, but they are within detection precision. New event 16 is coincident and displaced with the earlier label 713 TeV. New event 23 is not coincident with earlier label 190 TeV, but within error detection.
In the earlier (2014) maps we did pointed, following UHECR clustering events, UHE neutrino 206 TeV  and now the additional event 26 as tagged in Figures \ref{fig:3} and \ref{fig:5} by letter H (H1, H2);
we also labeled the region where it has been event 709, now as a letter C or by new event $10$. These additional event (26) confirm our proposal for a galactic UHE $\nu$ source. In the following table the UHE $\nu_{\mu}$ events in bold are due to recent (2016) New events trough-going muons from left to right (from large $\alpha$ angle to small ones considering $\delta$ up and down),  old trough-going muons (2015) label by their energy in TeV number. Each of the 12 coincident (old-new) event is tagged by a star symbol. The 8 clustering regions labeled by capital letters from ``A'' to ``I'' are connecting old and new pair events; The maximal deflection between old and new direction
is shown in last table and it is related to event $5$ and it is as large as $3.7^\circ$.
The spread angle for each pair is shown in figures by colored arrows  A-I and $\delta\theta$. Because of  the IceCube maximal error  the allowable correlation
opening angle considered is twice as large ($\leq 6.7^\circ$); for such a large maximal distance the arrow are red;
for the intermediate opening angles and for the most collimated ($\leq 3^\circ$) events the arrow turn to green.
The label ``del.'' means that this event was deleted with the new threshold.
The \textsl{shower} label indicates that around these correlated UHE muons there are shower cascade events.
Let us comment the table  containing the recent 2015-2016 trough going muons; first we could identify the
coincident UHE muon track by its date birthday. Second we comment from $\alpha = 360^\circ \to 0^\circ$ the event sequence;
the bold number tag the new (2016)  muon tracks while the normal size number tag the TeV energy of the old 2015 events.
There are 21 (2015) muons and additional 29 (2016) track events: most (13) are just the same
events updated as mentioned before, with some  change of energy and even of directionality.
For instance the old 2015 event of 1693 TeV is the same as the new event \textbf{6}, but its direction is deflected by nearly $2.2^{\circ}$ degree. The maximal deflection occurred for the new event \textbf{5} or old 178 TeV  by nearly $3.7^{\circ}$. This imply that the accuracy in IceCube maybe assumed not as sharp as $0.4^{\circ}$  but as  much  as $3.7^{\circ}$. This change the correlation pairs ability. The new possibility offer the connection between several couple of events up to twice $3.7^{\circ}$ as shown figure above, some of them bounded to UHECR narrow clustering.

\begin{table}[ht]
\centering
\label{table:1}
\vspace*{3pt}
\resizebox{1.05\columnwidth}{!}{%

\begin{tabular}{@{}rlcccr@{}}
\hline
Ev. & Jul. Day      & \multicolumn{2}{c}{Coordinates} & E (TeV)    &  Notes\\
          &                        & $\delta$  &  $\alpha$                 &                           & \\
 \hline
\textbf{3}       &       55355.5     &    23.8    &    344.93                  & 340   &  $\star$ A $\Delta\theta = 6.5^\circ$\\
442   &       55355.5     &       24   &      346.8                   & 442   &  $\star$ A $\Delta\theta = 6.5^\circ$ \\
\textbf{25}     &       56799.96   &   18.05  &      349.39                 & 400   & A $\Delta\theta = 6.5^\circ$\\
\textbf{8}       &       55478.38   &   11.09  &    331.08                   & 660   &  \textcolor[gray]{0.5}{$\star$}\\
466   &       55478.38   &    11      &      331                      & 446   &  \textcolor[gray]{0.5}{$\star$}\\
\textbf{18}     &       56146.21   &   1.57    &      330.1                   & 260   & B $\Delta\theta = 6.9^\circ$\\
193   &       55405.5     &   2.8      &      323.3                   & 193   &  del. B  $\Delta\theta = 6.9^\circ$\\
\textbf{14}    &        55764.22   &   5.29    &      315.66                 & 210   &  \\
\textbf{11}      &      55589.56   &   1.03    &    307.71                   & 240   &  \textcolor[gray]{0.7}{$\star$}\\
184   &       55589.6     &      1      &    307.9                    & 184   &  \textcolor[gray]{0.7}{$\star$}\\
178    &      55387.5     &   21.9    &    310.5                    & 178   & $\dag$ $\Delta\theta = 3.65^\circ$\\
\textbf{5}        &      55387.54   &   21.00   &    306.96                 & 230   &  $\dag$ $\Delta\theta = 3.65^\circ$ \\
\textbf{2}        &      55141.13   &   11.74   &    298.21                 & 250   &  \\
\textbf{n.33}       &      56221       &   7.8       &    292.5                   & 385   &  \textsl{shower} \\
709     &     55513.60   &   3.1       &    285.7                   & 709   & \textcolor[gray]{0.5}{$\dag$} C\\
\textbf{10}       &     55513.60   &   1.03     &    285.95                 & 520   &  \textcolor[gray]{0.5}{$\dag$} C\\
\textbf{24}       &     56666.5     &   32.82   &    293.29                 & 850   &  \\
191    &      55834.4     &   -4.4      &    267.6                  & 191    &  del.\\
\textbf{7}        &      55464.9    &   13.4      &    266.29                   & 460   &  \textcolor[gray]{0.7}{$\dag$}\\
400   &       55464.9    &      13.8    &    267.5                    & 400   &  \textcolor[gray]{0.7}{$\dag$}\\
\textbf{13}       &     55722.43   &   35.55   &    272.22                 & 210	   &  \\
198     &     55829.3    &   52.7      &    277.5                   & 198	   & del. \\
n.17       &     55804.37   &   14.5     &    247.4                   & 200   &  \textsl{shower} \\
1693   &       55421.5     &    16.3    &    254                   & 1693   &  $\ast$ $\Delta\theta = 2.2^\circ$ \\
\textbf{6}         &       55421.5     &   15.21   &   252.00               & 770   &  $\ast$ $\Delta\theta = 2.2^\circ$\\
\textbf{12}       &       55702.77   &   20.30  &    235.13               & 300   &  \textcolor[gray]{0.5}{$\ast$} D\\
393     &       55702.77   &    19.3    &    235.2                  & 393   &  \textcolor[gray]{0.5}{$\ast$} D $\Delta\theta = 3.2^\circ$\\
394       &     55987.8     &   18.9   &    238.3                 & 394   &  D $\Delta\theta = 3.2^\circ$\\
158      &      55896.86    &   3.2      &    221.9                & 158   & del. \textcolor[gray]{0.7}{$\ast$} E $\Delta\theta = 6.6^\circ$\\
\textbf{15}       &       55896.86    &   1.87    &    222.87              & 300   &  \textcolor[gray]{0.7}{$\ast$} E $\Delta\theta = 6.6^\circ$ \\
\textbf{22}       &       56521.83   &   -4.44   &    224.89                 & 400	   &  E $\Delta\theta = 6.6^\circ$ \\
\textbf{19}       &       56211.77   &   -2.39   &    205.11                 & 210	   &  \\
185       &       55642   &       6.7   &    207.2                   & 185	   & del. \\
\textbf{17}       &       56062.96   &   31.96   &    198.76                 & 200	   &  \\
\textbf{20}       &       56226.6   &   28.04   &    169.61                 & 750	   &  \\
189                &       55370.7   &   47.6   &    138.9                 & 189	   & $\ddag$ \\
\textbf{4}       &       55370.74   &   47.8   &    141.25                 & 260	   &   $\ddag$\\
156                &       55768.5   &   6.8   &    152.2                 & 156	   & del. \\
\textbf{27}       &       56819.2   &   11.42   &    110.63            & 4450	   &  \\
\textbf{26}       &       56817   &   1.29   &    106.26            & 340	   & F, H2  \\
\textbf{n.5}       &       55512.6   &   -0.4   &    110.6                 & 71.4	   &   \textcolor[gray]{0.5}{$\ddag$} H1 $\Delta\theta = 4.5^\circ$ \\
206               &       55512.6   &   0   &    110.5                 & 206	   &  del. \textcolor[gray]{0.5}{$\ddag$} H1 $\Delta\theta = 4.5^\circ$ \\
\textbf{28}       &       57049.48   &   4.56   &    100.48            & 210	   & F $\Delta\theta = 6.7^\circ$\\
\textbf{9}       &       55497.3   &   0.5   &    88.95                 & 950	   &   \textcolor[gray]{0.7}{$\ddag$}\\
880                &       55497.3   &   0.2   &    88.5                 & 880	   &  \textcolor[gray]{0.7}{$\ddag$} \\
\textbf{21}       &       56470.11   &   14.46   &    93.38                 & 670	   & G $\Delta\theta = 2.2^\circ$\\
\textbf{29}       &       57157.94   &   12.18   &    91.6                 & 240	   & G $\Delta\theta = 2.2^\circ$\\
\textbf{16}       &       55911.3   &   19.1   &    36.65                 & 660	   &   $\diamond$\\
713                &       55911.3   &   18.6   &    37.1                 & 713	   & $\diamond$ \\
\textbf{23}       &       56579.91   &   10.2   &    32.94                 & 390	   & I $\Delta\theta = 2.3^\circ$ \\
190               &       55803   &   11.8   &    31.2                 & 190	   & del. I $\Delta\theta = 2.3^\circ$ \\
\textbf{1}       &       55056.7     &    1.23    &    29.51                  & 480   & \\
179               &       55806.1   &   7.8   &    9.4                 & 179	   &  \\
\hline
\end{tabular} }
\caption{All the table note and description are in the text (in the subsection:\ref{Events in detail and Table reading})
}
\label{my-label}
\end{table}

\section{Probability and Conclusions}
Let us ask what it is the probability to find within the Fig. \ref{fig:3} a correlated event (C, H) in next 2015 and in final 2016 sample. For the three black arrow considered the angular area maybe take within a minimal size $4.5^\circ$ (containing C, H1, H2) or as large as $6.7^\circ$ containing (containing C, H1, H2 and F);
The sky area where UHE muon neutrino may shine and they may be detected is toward the North and its value is nearly $\simeq15.000^\circ$ because one must cut the North pole terrestrial opacity to 200 TeV neutrino.
Each of the three candidate source (label by the three arrows in Fig. \ref{fig:3}) assuming an opening angle of $6.7^\circ$ has a ratio area of being observed as large as $9 \cdot 10^{-3}$; the total solid angle is three times larger: $2.8 \cdot 10^{-2}$. The  binomial probability to discover within $\simeq 15.000^\circ$ on 2015 data of 21 trough going muons just two hit (C, H1) (and not discover 19 hit) in such a large test area is $P_{2015} =  9.5 \cdot 10^{-2}$.
Naturally the correct probability, assuming a more realistic narrow spread angle of $4.5^\circ$ (C, H1) is much rare,
  as small as $2.4 \cdot 10^{-2}$; however the next trial (2016) need a separation angle for event F as large as $6.7^\circ$. Consequently the probability to observe on 2016 an additional pair of event (event H2,F) require a wide angle
  as assumed in the beginning $6.7^\circ$. The probability to occur by chance for the new 29 events (12 are coincident to old events) is simply the case to find 2 hit in 17 trials: $P_{2016} =  7 \cdot 10^{-2}$.
   Therefore, the recent sample of trough going muons did reconfirm the validity of our assumption:
   UHECR narrow clustering may hint the underline UHE$\nu$ source.
   The overall probability is just the product of the two ones above:
   $$
   P_{tot} =  P_{2016} \cdot P_{2015} = 6.65\cdot 10^{-3}
   $$
   Thus, at the present we claim that the probability that a neutrino sources discover by UHECR clustering is not correct,
   is well below $1 \%$.
%
%

The most compelling absence of Virgo in UHECR sky forced us to UHECR light or lightest nuclei composition
\cite{6b}. Such UHECR He-Be-B bending is well compatible with the very smoothed signal  by twin Hot Spots
clustering. They probably might be correlated to nearest AGN, M82 in the North and Cen A in the South \cite{6, 19}.
Also LMC, SMC, NGC 253, Fornax D, may also shine UHECR \cite{19}. 
The extragalactic flight is so far and so long (by random walk in magnetic fields) that
the correlated UHE neutrino might be diluted and maybe hidden. On the contrary  very narrow clustering, as from Vela
 and clusters other ones may hint for nearby (mostly galactic)
sources whose arrival and timing maybe much shorter and they may be  still correlated to UHE neutrinos activity.
Because of it we suggested since 2013-2014 to look for the UHE trough going or muon crossing in IceCube and to correlate them with narrow UHECR clustering.
Among the area considered a clustering toward SS433 (event C) remains  a very promising one. Additional clustering (narrow doublet H1, H2 and a third, F) are here confirmed as a very probable source; new rare trough going muons along a chain of UHECR (event I) and the additional narrow clustering (G, D, E) seem to confirm the strategy narrow clustering UHE muon crossing event. The event B is not peculiar at all. However, the event A is surprising.

\section{UHECR by UHE \texorpdfstring{$\nu$}{n} scattering on relic \texorpdfstring{$\bar{\nu}$}{s} ?}
The UHE $\nu_{\mu}$ is well correlated to a train of UHECR events in A: indeed, it might be connected to the most powerful huge Fermi source in $\gamma$: AGN 3C 454, Fig.\ref{fig:2},\ref{fig:5}. However it is located half a way across the universe; in this view our earlier model (1997-1999)  able to overcome  far distance well above GZK bound \cite{Fargion:1997ft}, is  needed. It is, in fact, based on relic neutrino in dark hot halos offering the role of calorimeter  for UHE ZeV neutrinos whose scattering makes Z bosons. We already noted that such a far AGN may be an active source of UHECR if the UHE $\simeq 10^{21}$ eV  neutrino scattering on relic ones may produce Z resonance , : its ultra relativistic decay in flight may overcome of the severe GZK cut off. The result indeed if will be confirmed might be the first discover of the UHE neutrino scattering on the relic neutrino whose mass may range in an allowable mass of $0.4-0.1$ eV  \cite{Fargion:2006uz}, compatible with cosmic bounds. This discover might be the most spectacular astrophysical road-map to a neutrino mass detection and measure, as well as the best tool to reveal neutrino clustering halos. Because the IceCube muon neutrino sky in confined to the celestial North (the south is widely polluted by downward muons) there are a large number of sources that are not observable as: Vela, GC, SMC and LMC, Cen~A, Fornax~D and NGC~253. These sources might be observed via HESE event born inside IceCube.
The neutrino muon trough going tracks are probably, already pointing to few defined galactic sources as SS433, (C or H).
 The complementary $\tau$ air-shower (upward or horizontal) neutrino astronomy should (or must) be born too
\cite{0e1, 0e3, 0f}; trough-going muons \cite{0c} and tau airshowers, both will be the main currier of the most hidden and enigmatic UHE neutrino sky.



\bibliographystyle{elsarticle-num}
\bibliography{FargCRIS2016,farg_inspire_2016_10}


\end{document}